\documentclass[seceq,preprint]{ptptex}

\usepackage{amsmath,amsfonts,latexsym,amssymb}
\usepackage{verbatim,amsthm,graphicx}
\usepackage{wrapft}

\usepackage{amsmath}
\usepackage{amssymb}
\usepackage{mathrsfs}

\def\scrI{{\mathscr I}}

\def\scri{\scrI}

\def\defscript{\mathscr}

\def\B{{\defscript B}}

\def\H{{\defscript H}}
\def\I{{\defscript I}}

\def\M{{\defscript M}}

\def\ifempty#1{\def\tmpdata{#1}\ifx\tmpdata\empty }

\def\linebreak{\hfill\break}

\def\bra<#1|{\langle #1\rvert}
\def\ket|#1>{\lvert#1 \rangle}
\def\braket<#1|#2>{\langle #1|#2 \rangle}

\def\otop#1{\hbox{$#1\kern-0.1em$\llap{\hbox{\raise1.7ex\hbox{$\scriptstyle\circ$}}}} }

\def\inpare#1{\left(#1\right)}
\def\bigpare(#1){\left(#1\right)}

\def\insbra#1{\left[ #1 \right]}
\def\inang#1{\left\langle {#1} \right\rangle}

\def\bigbra[#1]{\left[ #1 \right]}

\def\cases#1{\left\{ \begin{array}{ll}#1\end{array}\right.}
\def\h{\hat }
\def\t{\tilde }

\def\cd{\cdot }
\def\b{\bar }

\def\tend{\rightarrow}

\def\equivalent{\quad\Leftrightarrow\quad}
\def\therefore{\mbox{\setbox0=\hbox{X}\hbox{$\ldotp$}\raise0.7\ht0\hbox{$\ldotp$}\hbox{$\ldotp$}} \quad }
\def\because{\mbox{\setbox0=\hbox{X}\raise0.7\ht0\hbox{$\ldotp$}\hbox{$\ldotp$}\raise0.7\ht0\hbox{$\ldotp$}}\kern0pt }

\def\SO{{\rm SO}}

\def\UG{{\rm U}}

\def\upin{\hbox{\setbox0=\hbox{$\cup$} \vrule width 0.05 \wd0 height \ht0 depth 0pt \kern - 0.5\wd0 \box0 }}
\def\Frac(#1/#2){\left(\frac{#1}{#2}\right)}

\def\sdprod{\mathrel{{\setbox0=\hbox{$\displaystyle\times$}\lower0.3\wd0\hbox{$\stackrel{\box0}{\scriptstyle\sim}$}}}}

\def\tosigma#1,{%
    \ifx\tmpindex\relax \def\tmpindex{#1} \let\next=\tosigma
    \else \ifnum\tmpindex=0 1 \else \sigma_\tmpindex \fi
          \ifx#1\relax  \let\next=\relax
          \else \otimes \let\next=\tosigma \def\tmpindex{#1} \fi
    \fi \next}
\def\tspb(#1){\let\tmpindex=\relax\tosigma#1,\relax,}
\def\Order#1{{\rm O}\!\left(#1\right)}

\def\pd{\partial}

\def\Eq#1{\begin{equation} #1 \end{equation}}

\def\Eqr#1{\begin{eqnarray} #1 \end{eqnarray}}
\def\Eqrn#1{\begin{eqnarray*} #1 \end{eqnarray*}}

\def\Eqrsubl#1#2{\begin{subequations}
  \expandafter\ifx\csname Rlabel\endcsname \relax \label{#1}
  \else \Rlabel{#1} \fi \Eqr{#2}\end{subequations}}
\def\Bitm{\begin{itemize}}
\def\Eitm{\end{itemize}}
\def\Blist#1#2{\begin{list}{#1}{\parsep=0pt \itemsep=0pt%
  \listparindent=0pt #2}}
\def\Elist{\end{list}}
\long\def\ignore#1#2{\def\ignoreflag{#1}\long\def\tmptext{#2}
  \ifnum\ignoreflag>1 #2 \fi}

\preprintnumber[3cm]{KEK-Cosmo-1}
\title{\large Superradiance and Instability of Black Holes}

\author{Hideo {\sc Kodama}%
\footnote{E-mail: Hideo.Kodama@kek.jp}
}
\inst{Cosmophysics Group, IPNS, KEK and the Graduate University of Advanced Studies,
1-1 Oho, Tsukuba 305-0801, Japan
}

\abst{
We discuss the relation between the superradiance phenomenon and the instability of rotating black holes in higher dimensions. In particular, we point out that the superradiant instability of a massless scalar field around a simply rotating Kerr-adS black hole implies the gravitational instability of that black hole for tensor-type perturbations.
}

\begin{document}
\maketitle

\section{Introduction}

The stability of black holes has a crucial importance when we study their formation and fate in Nature. It also has an intimate relation with the cosmic censorship hypothesis. For these reasons, this problem has been studied for a long time, and uniqueness and perturbative stability were established for most asymptotically flat black holes in the four-dimensional Einstein-Maxwell system except for the Kerr-Newman black hole\cite{Heusler.M1996B,Kodama.H2004a}.

When we go beyond this classical system, we encounter various new situations. One such extension is to consider systems containing other types of matter. Very interesting examples are the Einstein-Skyrme system and the Einstein-Yang-Mills system. These systems have three families of static asymptotically flat spherically symmetric solutions; a soliton family, a hairy black hole family and the vacuum one. These all families of solutions ware shown to be stable numerically for the Einstein-Skyrme system while for the EYM system, the colored black holes and the soliton solutions were shown to be unstable (see Ref. \citen{Volkov.M&Galtsov1999} for review). Hence, in the latter case, the uniqueness theorem holds practically.

Another extension motivated by recent progresses in unified theories is to consider higher-dimensional black holes. In the static AF Einstein-Maxwell system, the uniqueness theorem still holds in higher dimensions. Further, in the vacuum case, the Schwarzschild-Tangherlini solution was shown to be stable\cite{Ishibashi.A&Kodama2003} by using the extension of the Regge-Wheeler and Zerilli equations to higher dimensions\cite{Kodama.H&Ishibashi2003}, although the stability of the charged static black hole  was proved only for $D=5$ analytically\cite{Kodama.H&Ishibashi2004} and for $6\le D\le 11$ numerically\cite{Konoplya.R&Zhidenko2007}.

If we consider non-asymptotically flat systems, however, the situation changes. The most notorious is the Gregory-Laflamme instability of  black string and branes\cite{Gregory.R&Laflamme1993,Seahra.S&Clarkson&Maartens2005,Kudoh.H2006}. A Gregory-Laflamme type instability is also predicted to occur around the rotation axis of rapidly rotating Myers-Perry black holes\cite{Emparan.R&Myers2003}(cf. Ref.\citen{Kunduri.H&Lucietti&Reall2006})

Non-vanishing cosmological constant also introduces subtleties. In four dimensions ($D=4$), static adS/dS black holes and their charged extensions can be shown to be perturbatively stable\cite{Ishibashi.A&Kodama2003,Kodama.H&Ishibashi2004}. The perturbative stability can be also proved for static dS black holes in $D=5$ and $6$ analytically\cite{Ishibashi.A&Kodama2003} and $7\le D\le 11$ numerically\cite{Konoplya.R&Zhidenko2007} for the neutral case and in $D=5$ analytically\cite{Kodama.H&Ishibashi2004} and $6\le D\le 11$ numerically\cite{Konoplya.R&Zhidenko2007} for the charged case. This suggests that the introduction of a positive cosmological constant does not affect the stability property of a black hole. In contrast, the stability property of adS black holes is not certain in $D\ge5$, except for topological black holes with zero or negative mass\cite{Birmingham.D&Mokhtari2007A}

Finally, we encounter surprising new and rich phenomena when we study rotating black holes in higher dimensions. In particular, the discoveries of various black ring solutions\cite{Emparan.R&Reall2006} has revealed the fact that rotating black holes are far from unique in five dimensions and possibly in other higher dimensions. This fact should, however, be taken with care because only stable solutions will be realised in Nature, at least as final states. In fact, many people seem to suspect that black ring solutions would suffer from a Gregory-Laflamme type instability. 

Unfortunately, however, no exact analysis of perturbative stability has been done for most rotating black objects in higher dimensions so far. One exception is the Kerr-adS black hole. In four and five dimensions, this black hole were conjectured to be stable for slow rotation $|a| \ell< r_h^2$ and unstable for rapid rotation violating this condition\cite{Hawking.S&Reall1999}, where $a$ is the angular momentum parameter, $r_h$ is the horizon radius and $\ell$ is the curvature radius of the asymptotic adS spacetime. In four dimensions, this conjecture was proved to hold numerical in some limiting cases by Cardoso, Dias and Yoshida\cite{Cardoso.V&Dias2004,Cardoso.V&Dias&Yoshida2006}. 

This instability is understood to be caused by superradiance. As is well-known, the superradiance is the phenomenon that the amplitude of a bosonic field wave incidental to a Kerr black hole is amplified after reflection for some range of frequencies\cite{Chandrasekhar.S1983B}. Press and Teukolsky\cite{Press.W&Teukolsky1972} pointed out that this superradiance provokes an instability if a Kerr black hole is surrounded by a reflective mirror wall\cite{Zeldovich.Y1971,Cardoso.V&&2004}. Later on, it was suggested that this superradiance instability would  occur even without such a wall if we consider a massive scalar field, because the barrier of its effective potential plays the role of the mirror\cite{Damour.T&Deruelle&Ruffini1976}, although the growth rate of this instability is too small to be effective in the real world except for the black hole of the Planck mass scale\cite{Zouros.T&Eardley1979,Detweiler.S1980,Furuhashi.H&Nambu2004,Dolan.S2007A}. This mechanics was used to explain the instability of a five-dimensional spinning black string solution\cite{Marolf.D&Palmer2004,Cardoso.V&Yoshida2005} and to predict the instability of large doubly rotating black rings\cite{Dias.O2006}. 

The most peculiar feature of asymptotically adS spacetimes is its global non-hyperbolicity. This is caused by the fact that the spatial infinity is causally at a finite distance and forms a time-like boundary. Due to this feature, a scalar field around  a rotating black hole in an adS spacetime may suffer from a superradiance instability even the field is massless. This is the instability pointed as by Hawking and Reall. 

In this paper, we show that we can utilise this argument on the superradiance instability of a massless scalar field to study the gravitational stability of Kerr-adS black holes.

\section{Superradiance of Asymptotically Flat Black Holes}

We first outline the derivation of the superradiance condition for asymptotically flat rotating black holes in a form that does not depend on spacetime dimension or horizon topology. 

\subsection{Symmetry and Asymptotic Structure of Spacetime}

\begin{figure}[t]
\centerline{\includegraphics[height=5cm]{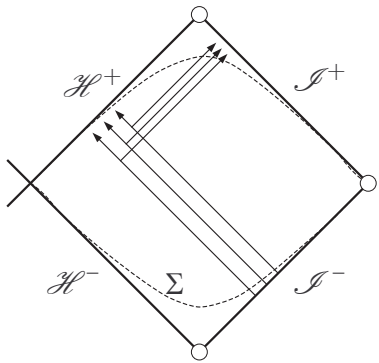}}
\caption{Global structure of DOC of an asymptotically flat black hole}
\label{fig:AFBH}
\end{figure}

For an asymptotically flat black hole spacetime, the DOC $\M$ is simply connected and globally hyperbolic, and its boundary consists of four components corresponding to the future and past infinities and the future and past horizons\cite{Hawking.S&Ellis1973B,Chrusciel.P&Wald1994,Galloway.G1995}:
${
\partial \M = \H^+ \cup \H^- \cup \scri^+\cup \scri^-
}$ (see Fig.\ref{fig:AFBH}).

In the present paper, we only consider a stationary rotating black hole with analytic metric. Then, from the rigidity theorem\cite{Holland.S&Ishibashi&Wald2007}, there exist commuting rotational Killing vectors $\eta_i$ ($i=1,\cdots,N$) with $N\ge1$ in addition to the time translation Killing vector $\xi$. Let $(y^p)=(t,\varphi^i)$ be coordinates in orbits of $(\xi_p)=(\xi,\eta_i)$ such that $\xi_p y^q=\delta_p^q$, and $x^a$ be coordinates labeling orbits such that $\xi_p x^a=0$. Then, the spacetime metric can be written
\Eq{
ds^2=g_{pq}(x) \chi^p \chi^q
   + q_{ab}(x) dx^a dx^b,
}
where $\chi^p$ is a 1-form written as
$
\chi^p = dy^p + A^p_a(x) dx^a.
$
%

At infinity, this metric approaches the Minkowski spacetime and has the asymptotic form
\Eq{
ds^2= -f(r)dt^2  + \frac{dr^2}{f(r)} + r^2 d\Omega_{D-2}^2
         + \Order{\frac{1}{r^{D-2}}},
}
where $D$ is the spacetime dimension, $d\Omega_n^2$ is the metric of unit Euclidean $n$-sphere $S^n$, and $f(r)$ is 
$
f(r)= 1 -{2 M}/{{r}^{D-3}}. 
$
%

In contrast, the spacetime is largely deformed due to rotation near the horizon. In higher dimensions, the rotation is characterised by the (multi-component) angular velocity  $\Omega_h^i$ of the horizon, which is defined through the expression $k=\xi + \Omega_h^i \eta_i$ for each null geodesic generator $k$ of the horizon $\H^+\cup\H^-$ in terms of the Killing vectors. This angular velocity can be determined from the metric in the following way. First, because $k$ is a null vector, the angular velocity of the horizon $\Omega_h^i$ satisfies 
${
g_{tt} + 2g_{t i} \Omega_h^i + g_{ij} \Omega_h^i\Omega_h^j=0
}$.
Since $k$ is the unique null direction in the $(N+1)$-plane spanned by $\xi$ and $\eta_i$, this equation must have a single solution for $(\Omega_h^i)$ as a vector. Hence, $\Omega_h^i$ can be expressed as
${
\Omega_h^i =\Omega|^i_{\text{horizon}} \ (
\Omega^i:=-g^{ij} g_{tj})
}$,
and the horizon location is determined by 
${
\b\Delta := -g_{tt}+g^{ij}g_{ti}g_{tj}=0
}$.

In order to define a regular coordinate system around the horizon, we introduce a new variable $\t\varphi^i$ that is constant along the null geodesic generators:
${
k\t\varphi^i=0,\quad \eta_j \t\varphi^i=\delta^i_j.
}$
These conditions determine $\t\varphi^i$ as
${
\t\varphi^i= \varphi^i - \Omega_h^i t
}$,
upto a constant independent of $t$ and $\varphi^i$. By introducing an appropriate function $r$ which is equal to $r_h$ on the horizon, we can rewrite the metric near the horizon as
\Eqr{
ds^2 &=& -\frac{\Sigma^2\Delta}{\Gamma}(\chi^t)^2
   + g_{ij}\insbra{\tilde\chi^i-(\Omega^i-\Omega^i_h)\chi^t}
      \insbra{\tilde\chi^j -(\Omega^j-\Omega^j_h)\chi^t}
    \notag\\
    && + \frac{\Sigma^2}{\Delta}dr^2 + q_{AB}(r,z) dz^A dz^B,
\label{metric:nearhorizon}
}
where $\t\chi^i=d\t\varphi^i+A^i-\Omega^i_h A^t$, $\Delta$ is a function only of $r$, and $\Sigma^2$ and $\Gamma$ are regular functions around the horizon such that
${
\b \Delta= {\Sigma^2}\Delta/{\Gamma}
}$ and
${g_{rr}= {\Sigma^2}/{\Delta}
}$.
We can show  that $\Gamma$ is equal to a constant $\Gamma_0$ on the horizon from the zeroth-law of black holes because it is related to the surface gravity $\kappa$ of the black hole by 
${
2\kappa = \Gamma_0^{-1/2}\Delta'(r_h)
}$. 
Then, in terms of the coordinates $(u_+,\t\varphi,r,z^i)$ where 
${
u_+ = t + r_*
}$
with 
${
r_*:=\int^r \Gamma_0^{1/2}{dr}/{\Delta}
}$,
we can put the metric in a form that is regular around the future horizon. Similarly, if we use $u_-=t-r_*$ in place of $u_+$, we obtain a regular coordinate system around the past horizon.

\subsection{Scalar fields}

Let us consider a massless free scalar field with charge $q$ that satisfies the Klein-Gordon equation
${
D^\mu D_\mu \phi=0
}$
with 
${
D_\mu = \pd_\mu - iq A_\mu.
}$
For any complex solutions $\phi_1$ and $\phi_2$, the Klein-Gordon product defined by
\Eq{
I(\phi_1,\phi_2):=- i \int_\Sigma \insbra{\b\phi_1 D^\mu\phi_2
 -(\b D^\mu \b\phi_1)\phi_2} \, d\Sigma_\mu
}
is independent of the Cauchy surface $\Sigma$ in a global hyperbolic domain of the spacetime. 

\subsubsection{Asymptotic behavior at infinity}

From the isometry, we can expand the massless scalar field $\phi$ into eigenmodes with respect to $\xi$ and $\eta$ as
${
\phi= \h\phi(r,z)e^{-i\omega t + im\cdot \varphi}
}$,
where $m\cdot\varphi=m_i\varphi^i$ with $m_i$ being a set of integers. The amplitude of each eigenmode $\h \phi$ behaves at infinity as
\Eq{
\h\phi \approx \frac{1}{r^{D/2-1}}\inpare{ A e^{-i\omega r_*} + B e^{+i\omega r_*}},
}
where $A$ and $B$ are bounded functions of $z^i$, and $r_*$ is defined by
${
d r_* = {dr}/{f(r)}
}$.
For non-rotating spherical black holes, this function is identical to $r_*$ introduced above near the horizon, but they have no direct relation in general.

\subsubsection{Boundary condition at the horizon}

Near the future (or past) horizon, the coordinates $(u_+,\t\varphi^i,r,z^A)$ are regular. In terms of these coordinates, the eigenmode is written
\Eq{
\phi=\h \phi(r,z) e^{i\omega_* r_*} e^{-i\omega_* u_+ + im\cdot \t\varphi},
}
where
${
\omega_* = \omega - m_i \Omega_h^i
}$.
Here, from the asymptotic behavior of the equation for $\h\phi$ near the horizon, $\h\phi$ behaves as 
${
\h\phi(r,z) \approx C(z) e^{-i\omega_* r_*}+ D(z) e^{+i\omega_* r_*}
}$
near the horizon. This becomes purely ingoing at horizon if $D=0$ and 
\Eq{
\h\phi(r,z) \approx C(z) e^{-i\omega_* r_*}.
\label{IngoingCondition:horizon}}
%

\subsubsection{Superradiance}

Let us consider Cauchy surfaces $\Sigma$ that extend to spatial infinity and have the common inner boundary at the bifurcating $n$-manifold $\H^+\cap\H^-$. If we take the past limit of such surfaces, they approaches $\Sigma_-=\H^-\cup \scri^-$. Hence, assuming that there exists no bounded state around the black hole, the KG product in this limit can be simply given by the contribution from $\scri^-$,
\Eq{
I_{\scri^-} = i \int dv \int_{S^n} d\Omega_n \lim_{r\tend \infty} r^n (\b\phi\overset{\leftrightarrow}\pd{}_v \phi)
=\sum_m \int d\omega\,4\pi \omega \langle |A_{\omega,m}|^2\rangle,
}
where $v=t+r_*$ and
${
\langle |A|^2\rangle 
:= 2\pi \int_{S^n/S^1}d^{n-1}z\sqrt{\rho q}\, |A|^2
}$.

Next, let us consider the future limit of the Cauchy surfaces. In this limit, by a similar argument, we can assume that the field has support on $\Sigma_+=\H^+\cup \scri^+$ and its asymptotic behaviour is
\Eq{
\phi \approx \cases{
   \frac{B}{r^{n/2}} e^{-i\omega(t-r_*)} & \text{at\ }\scri^+,\\
   C e^{-i\omega_* u_+ + i m \cd \t\varphi} &\text{at\ }\H^+.
   }
}
The contribution from $\scri^+$ to the KG product, $I_{\scri^+}$, is given by replacing $A$ by $B$ in the above expression for $\Sigma_-$. 
To obtain $\I_{\Sigma_+}$, we have to add to this the contribution from $\H^+$ given by
\Eqr{
I_{\H^+} &=& i \int du_+ \int_{\B}d\t\varphi d^{n-1}z\, 
  \sqrt{\rho q} 
 \insbra{\b\phi D_{+} \phi -(\b D_{+}\b\phi)\phi }_{r=r_h}
 \notag\\
 &=& \sum_m \int d\omega \,4\pi (\omega_*-q\Phi_h) \inang{|C_{\omega,m}|^2}_\B,
}
where $\B=\H^+\cap \H^-$,
${
\inang{|C|^2}_\B :=  2\pi \int_{\B/S^1}d^{n-1}z\sqrt{\rho q}\, |C|^2,
}$
and $\Phi_h$ is the electric potential of the horizon with respect to infinity. 

Since the KG product is independent of the choice of a Cauchy surface with the fixed boundary, we obtain
${
I_{\scri^-}= I_{\scri^+} + I_{\H^+}
}$,
which generally implies
\Eq{ 
\omega \langle|A_{\omega,m}|^2\rangle
= \omega \langle|B_{\omega,m}|^2\rangle
  + (\omega_*-q\Phi_h) \langle|C_{\omega,m}|^2\rangle_\B
}
Hence, the transmission and reflection coefficients for the free massless scalar field are given by
\Eq{
1= T_{\omega,m}+ R_{\omega,m};\quad
T_{\omega,m}= \frac{\omega_*-q\Phi_h}{\omega} \frac{\langle|C_{\omega,m}|^2\rangle_\B}{ \langle|A_{\omega,m}|^2\rangle},\quad
R_{\omega,m}= \frac{\langle|B_{\omega,m}|^2\rangle}{ \langle|A_{\omega,m}|^2\rangle}.
}
From this, it follows that the reflection coefficient exceeds unity for modes satisfying the condition
\Eq{
\omega-m \cdot \Omega_h -q \Phi_h<0.
\label{SuperradianceCond:AFBH}
}

Here, note that in the WKB approximation, the field equation for $\phi = A e^{iS}$ yields $p_\mu p^\mu=0$, where $p^\mu$ is the kinetic momentum define by $p_\mu=\pd_\mu S -q A_\mu$. In terms of this kinetic momentum, the above superradiance condition can be written $k^\mu p_\mu>0$, where $k^\mu$ is the null generator of the horizon. This implies that $p^\mu$ is null, but is not future-directed. Hence, the superradiance occurs if there exists a particle state that is physically allowed at infinity but not at horizon. This is intuitively natural because under this condition, the particle with the negative energy $-p^\mu$ becomes physical at horizon and can be absorbed by the black hole. In particular, the superradiance by a black hole occurs for massless scalar fields if the black hole is rotating, irrespective of the spacetime dimensions or the horizon topology.

\begin{figure}
\centerline{\includegraphics[height=5cm]{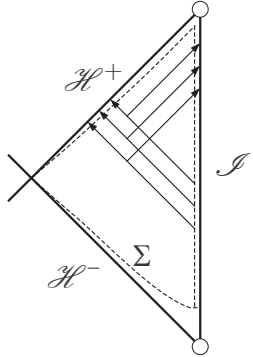}}
\caption{Global structure of DOC of an adS black hole}
\label{fig:AdSBH}
\end{figure}

\section{AdS-Kerr Black Holes}

\subsection{General Features}

AdS-Kerr black hole spacetimes have a couple of features that are quite different from those of asymptotically flat black hole spacetimes. Firstly, the DOC is not globally hyperbolic and its boundary consist of the past and future horizons and the spatial infinity:
${
\pd\M = \H^+ \cup \H^- \cup \scri
}$.
Second, although the spacetime can have the same symmetry structure as the asymptotically flat black holes, the time translation Killing vector $\xi$ is not unique in the asymptotically adS case, in contrast to the AF case.  This is because the norm of the static time-like Killing $\pd_t$ of the exact adS spacetime diverges in proportion to $r^2$ at infinity, while the norm of $\eta_i$ behaves in the same way. Hence, in a rotational spacetime, any linear combination $\xi+c^i\eta_i$ has the equal right as the Killing vector to define the time translation.
This freedom introduces an ambiguity in the definition of the angular velocity of a black hole given above. Apart from this point, the structure around horizon for an adS black hole is the same as that for an AF black hole.

\subsection{Superradiance?}

The features of an asymptotically adS black hole spacetime pointed out above make the argument of superradiance quite delicate. First, if we consider a hypersurface $\Sigma$ shown by the dashed line in Fig.\ref{fig:AdSBH} and take the limit such that $\Sigma$ approach the boundary of the DOC, the flux conservation law for a scalar field can be written
\Eq{
I_{\H^+}+ I_{\scri}= I_{\H^-}=0.
}

In order to estimate $I_{\scri}$, we need to know the asymptotic behaviour of the field. For a massless scalar field satisfying the KG equation, the leading part of the asymptotic behaviour does not depend on the properties of a black hole at the center and is the same as that in the exact adS spacetime. For $\phi\propto e^{-i\omega t}$, it is given by 
${
\phi \approx E_0 + {E_1}/{r^{D-1}}
}$.
Hence, if we require that the field configuration has a finite energy, we have to select the boundary condition $\phi \sim 1/r^{D-1}$ at $r\sim\infty$. For this boundary condition, we have
\Eq{
I_\scri= -i \int dt \int d^{D-2}\Omega
\lim_{r\tend\infty} r^{D-2}(1+r^2/\ell^2) \b\phi
 \overset{\leftrightarrow}{\pd}_r\phi
=0.
}
Hence, we are left with the funny condition
\Eq{
(\omega- m\Omega_h-q\Phi) \inang{|C_{\omega,m}^2|}_{\B/T^N} =0.
}

Clearly, the condition obtained above does not seem to be giving any information on superradiance. In fact, if we repeat the same argument for a rotating black hole inside a mirror box, we would obtain the same result. In this black hole bomb case, however, we know that instability occurs for modes with frequencies satisfying the standard superradiance condition \eqref{SuperradianceCond:AFBH}\cite{Cardoso.V&&2004}. Since the flux conservation law also holds in this case with instability, what is wrong with the above argument is the neglect of the contribution from the volume integral at finite distances. In the black hole bomb case, the energy extracted from the black hole is deposited in a region between the black hole and the mirror, and the contribution to the KG product from that region grows with instability. In the adS black hole case, a similar phenomenon may happen.

\subsection{Globally Timelike Killing Vector}

Hawking and Reall pointed out that the freedom in the choice of a timelike Killing vector has a very significant implication to the stability problem of an adS-Kerr black hole and the relevance of superradiance to it. Their argument goes as follows. First, if there is a Killing vector $\xi$ that is timelike everywhere outside the horizon, one can construct the conserved energy integral
\Eq{
\pd_t \int_{\Sigma(t)} d\Sigma n^\mu \xi^\nu T_{\mu\nu} =0,
}
where $n^\mu$ is the normal vector to the Cauchy surface $\Sigma(t)$. Here, if the matter satisfies the dominant energy condition, and $\xi$ is timelike everywhere, the integrant $n^\mu \xi^\nu T_{\mu\nu}$ is non-negative everywhere. Hence, any instability of the exponential growth type cannot occur.

In a stationary black hole spacetime, any Killing vector is spacelike or null on horizon because all Killing vectors are tangential to the horizon. Hence, a Killing vector that is timelike outside the horizon, if it exists, should be identical to the Killing vector $k$ that is parallel to the null geodesic generator of the horizon. Therefore, the problem is when $k$ is timelike everywhere outside the horizon. 

For example, the Kerr-adS solution in arbitrary dimensions was recently found by Gibbons, L{\"u}, Page and Pope\cite{Gibbons.G&&2005}. This solution is parametrised by the mass $M$, the multi-component angular momentum $a_1,\cdots, a_{N'}$ and the cosmological constant parameter $\lambda=-1/\ell^2$, where $N'=[(D+1)/2]-1$.  It is easy to show that $k$ is timelike at infinity if and only if 
\Eq{
\ell^2 \Omega_h^i{}^2 \le 1
\equivalent 
\ell^2 a_i^2 \le r_h^4 \ (i=1,\cdots,N').
\label{HRstabilityCondtion}
}
However, it is in general difficult to see whether $k\cdot k<0$ everywhere outside the horizon under this condition, 

Some exceptions are the case of the five-dimensional Kerr-adS solution discussed by Hawking and Reall\cite{Hawking.S&Reall1999} and the case in which the angular momentum components vanishes except for one component, say $a_1=a,a_2=\cdots=a_{N'}=0$. At least in these special cases, matter fields satisfying the dominant energy condition are stable if the condition \eqref{HRstabilityCondtion} holds, from the above argument by Hawking and Reall. Of course, this does not immediately implies that the black hole is gravitationally stable as well. 

\section{Instability of AdS-Kerr Black Holes}

A reduction of the Einstein equations for gravitational perturbations to a Teukolsky type equation is possible for an adS-Kerr black hole in four dimensions\cite{Giammatteo.M&Moss2005}. With the help of this formulation, it was shown that the four-dimensional adS-Kerr black holes are really unstable at least in the limit $r_h/\ell\ll1$ for modes satisfying the superradiance condition $\omega-m\Omega_h<0$\cite{Cardoso.V&Dias&Yoshida2006}. Although it is not imposed explicitly , the condition $\ell\Omega_h>1$ is practically satisfied under the superradiance condition because the real part of the eigen frequency $\omega$ for unstable modes are discrete and its minimum value is approximately $(m+2)/\ell$ for $m>0$. 

In dimensions greater than 4, however, no such reduction is available for gravitational perturbations of adS-Kerr black holes. One exception is the simply rotating black hole with $a_1=a,a_2=\cdots =a_{N'}=0$. The metric for such a black hole has $\UG(1)\times \SO(n+1)$ symmetry with $n=D-4$ in addition to the time translation invariance and can be written as
\Eq{
ds^2=g_{ab}(x) dx^a dx^b + S(x)^2 d\Omega_n^2,
}
where $d\Omega_n^2$ is the metric of the Euclidean sphere $S^n$. Hence, we can classify perturbations into scalar, vector and tensor types according to their transformation properties as tensors on $S^n$\cite{Kodama.H&Ishibashi&Seto2000}. In particular, for $n\ge3$, i.e., for $D\ge 7$, non-trivial tensor perturbations can exist, and the expansion coefficient $H_T$ of their amplitudes with respect to the tensor harmonics satisfies the hyperbolic equation on the four-dimensional spacetime with the metric $ds_4^2=g_{ab}(x)dx^a dx^b$,
\Eq{
\Box H_T - \frac{n}{S}DS\cdot D H_T + \frac{l(l+n-1)}{S^2} H_T=0,
\label{HTeq}
}
where $D_a$ is the covariant derivative with respect to the metric $ds_4^2$, $\Box=D^a D_a$ and $l$ is an integer greater than 1 labeling the tensor harmonics. 

\begin{figure}
\centerline{\includegraphics[height=6cm]{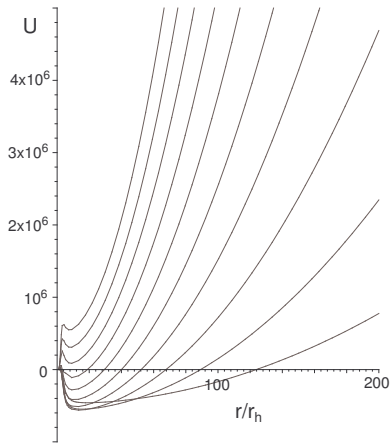}}
\caption{Plots of the effective potential $U$ for $D=7, \ell^2 a^2/r_h^4=0.9\sim 26, a^2/\ell^2=0.99, m=2\times 10^4, x=-0.99$.}
\label{fig:TPpot}
\end{figure}

A very interesting feature of this equation is that it is identical to the equation for spherical harmonic expansion coefficients of a massless scalar field in the same spacetime. Thus, the stability issue of the black hole for tensor perturbations can be reduced to the same issue for a free massless scalar field. In particular, from Hawking and Reall's argument, we can immediately conclude that the black hole is stable for tensor perturbations if $\ell^2 a^2 \le r_h^4$.

We can directly check this by writing the energy integral explicitly. In the coordinates in which the metric can be written
\Eqr{
ds^2 &=& -\frac{\Delta_r}{\rho^2}\inpare{dt-\frac{a}{C}\sin^2\theta d\phi}^2
          +\frac{\Delta_\theta\sin^2\theta}{\rho^2}\inpare{adt-\frac{r^2+a^2}{C}d\phi}^2
          \notag\\
   && + \frac{\rho^2}{\Delta_r}dr^2+ \frac{\rho^2}{\Delta_\theta}d\theta^2
   +r^2\cos^2\theta d\Omega_n^2,
}
the following energy integral is conserved:
\Eqrn{
\H(H_T) &:=& \int_{r_h}^\infty dr \int_{-1}^1 dx
  \left[ \frac{\Delta_r}{r^2+a^2} |\pd_r\Phi|^2
    +2\frac{(1-x^2)(2+\lambda a^2+\lambda a^2x)}{r^2+a^2}
    |\pd_x \Phi|^2
    \right.\notag\\
  && \quad \left.
  +\frac{F}{(r^2+a^2)\Delta_r} |\pd_t \Phi|^2
  +\frac{m^2 U_1+U_0}{r^2+a^2}|\Phi|^2 \right],
}
where $x=\cos(2\theta)$,
${
H_T= r^{-n/2} (r^2+a^2)^{-1/2} (1+x)^{-(n-1)/4} \Phi(t,r,x) e^{im\t\varphi}
}$,
and $F>0$.

We can show that both $U_0$ and $U_1$ are positive definite outside the horizon if $\ell^2 a^2 < r_h^4$. In this case, then, all terms in the energy integral is positive definite, and no exponential growth of $\Phi$ is possible. In contrast, for $\ell^2 a^2 > r_h^4$, $U_1$ becomes negative in some region. Hence, the effective potential 
${
U = \Delta ( U_0 + m^2 U_1)/F
}$
becomes negative in the same region for a sufficiently large value of $m$, as is illustrated in Fig.\ref{fig:TPpot}. Since $m$ does not appear inside $U_0$ or $U_1$ or in other places in the energy integral, this negative dip of the potential becomes deeper and deeper without bound as $m$ increases. Hence, it is quite likely that this type of black hole is unstable for $\ell^2 a^2 > r_h^4$ as in the four-dimensional case.

A similar result was obtained for the special Kerr-adS black hole with $a_1=\cdots=a_N=a$ in $2N+1$ dimensional spacetime by Kunduri, Lucietti and Reall\cite{Kunduri.H&Lucietti&Reall2006}.

\section{Discussion}

In this paper, we have pointed out that for simply rotating Kerr-adS black holes, its stability for tensor-type perturbations is equivalent to that of a massless scalar field around these black holes. From this it follows that those black holes are stable for tensor perturbations for slow rotation $\ell^2 a^2\le r_h^4$. Further, it also strongly suggests that they are unstable due to superradiance for rapidly rotating case. It will be interesting to check this by numerical calculations utilising the fact that the scalar field equation is separable in these background. It will be also interesting to calculate the instability growth time and see whether it blows up in the vanishing cosmological constant limit.

\section*{Acknowledgements}

The author thanks Akihiro Ishibashi for useful discussions. This work is partly supported by Grants-in-Aid for Scientific Research from JSPS (No. 18540265).



\end{document}
